\documentclass[10pt,twocolumn,twoside]{IEEEtran}

\usepackage{amsmath,amssymb,graphicx,caption,subcaption,float,multirow,adjustbox,authblk}
\usepackage[hidelinks]{hyperref}
\usepackage[table,xcdraw]{xcolor}
\newcommand{\subparagraph}{}
\usepackage{titlesec}

\titlespacing{\section}{4pt}{8pt plus 4pt minus 2pt}{0pt plus 2pt minus 2pt}
\titlespacing{\subsection}{2pt}{\parskip}{-\parskip}

\newcommand\numberthis{\addtocounter{equation}{1}\tag{\theequation}}

\title{Fast Switch Scanning Keyboards: Minimal Expected Query Decision Trees}

\author{\IEEEauthorblockN{Matt Higger$^*$\thanks{This work was supported by: NSF CNS-1136027, IIS1149570; NIH 2R01DC009834-06A1; NIDRR H133E140026.  \\ A complete package containing code and data associated with this document can be found online at \url{http://hdl.handle.net/2047/d20194049}.  Decision trees graphs were generated with \cite{TinevezTree}.},
Mohammad Moghadamfalahi, 
Fernando Quivira and
Deniz Erdogmus} \\
\IEEEauthorblockA{Electrical and Computer Engineering,
Northeastern University\\
360 Huntington Ave, Boston, MA 02115 \\
Email: higger@ece.neu.edu,
moghadam@ece.neu.edu,
quivira@ece.neu.edu,
erdogmus@ece.neu.edu}}

\begin{document}
\maketitle

\begin{abstract}
Augmentative and Alternative Communication (AAC) systems allow people with disabilities to provide input to devices which empower them to more fully interact with their environment.  Within AAC, switch scanning is a common paradigm for spelling where a set of characters is highlighted and the user is queried as to whether their target character is in the highlighted set.  These queries are used to traverse a decision tree which successively prunes away characters until only a single one remains (the estimate).  This work seeks a decision tree which requires the fewest expected queries per decision sequence (EQPD).  In particular, we remove the constraint that the decision tree needs to be a row-item or group-row-item style tree and minimize EQPD.  We pose the problem as a Huffman code with variable, integer cost and solve it with a mild extension of Golin's method in ``A dynamic programming algorithm for constructing optimal prefix-free codes with unequal letter costs", IEEE Transactions on Information Theory (1998).  Additionally, we model the user on the query level by their probability of detection and false alarm to derive their expected performance on the character level given some decision tree.  We perform experiments which show that the min EQPD decision tree (Karp) may reduce selection times, especially for \textit{timed} (single switch) switch scanning.
\end{abstract}

\begin{IEEEkeywords} 
AAC, switch scanning, row-column, block-row-column, Huffman, Karp
\end{IEEEkeywords}

\section{Introduction}\label{sec:intro}
In a switch scanning Augmentative and Assistive Communication (AAC) spelling system a subset of characters is highlighted.  The user indicates their desired character by telling the system whether their target character is in the highlighted query set \cite{Foulds1975}.  As an illustrative example, consider the alphabetical linear switch scanning spelling task.  Such a system queries the user alphabetically (`Is your target letter A?',  `Is your target letter `B?' ...).  Under this scheme it would take 1 query to select `A', 2 queries to select `B' and so on.  Letters earlier in the alphabet will require fewer queries for selection.  We could lower the expected queries per character decision (EQPD) by querying letters in decreasing prior probability.  This improved system would spell faster by spending its quickest slots in the query order on the most common characters.  

\begin{figure*}
\centering
\includegraphics[width=0.8\textwidth]{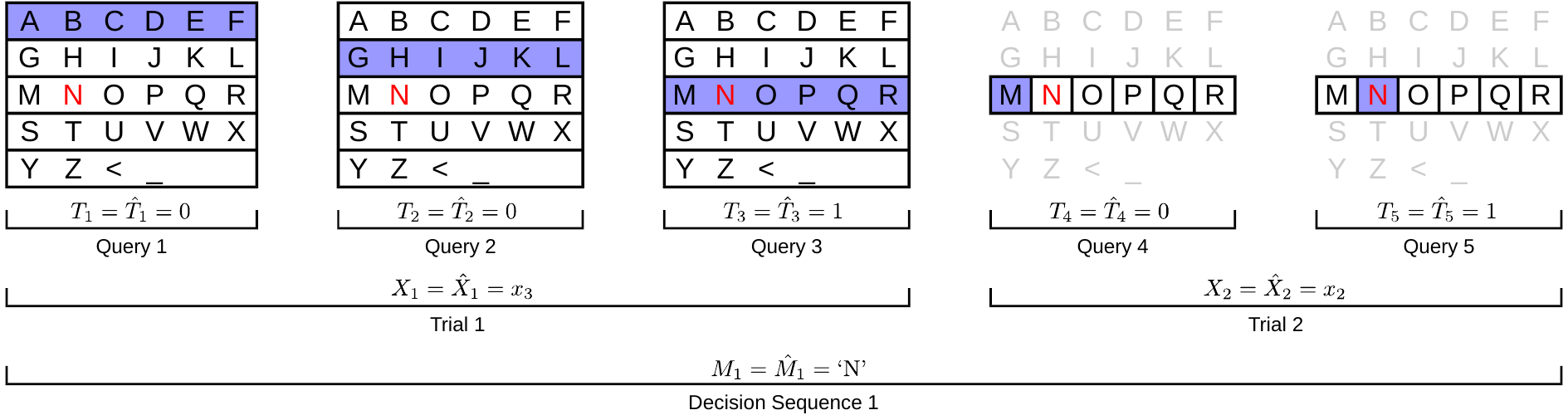}
\caption{Alphabetical $5 \times 6$ row-item decision sequence with no users errors.  The user produces a response only during the queries where their target character, `N' in this case, is in the blue highlighted query set.  The codeword $c_i = [x_3, x_2]$.  Of course the target character is red for illustrative purposes only; the system does not know the target character a priori.  Please see Sec \ref{sec:notationAndVocab} for notation details. \label{fig:fig1}}
\end{figure*}

We can further improve the system by first querying the user on subsets of letters, `Is your target letter in `A, B, C, D?', to hone in on a particular letter.  Many AAC systems adopt this proper-subset style query in row-item\footnote{We use the term `row-item' rather than `row-column'; they are both used in the literature but we feel `item' is more intuitive as a full column is not often queried.} or block-row-item scanning \cite{Abascal2004, Bhattacharya2008, Felzer2009}.  In a row-item scheme (see example in Fig \ref{fig:fig1}) the characters are arranged in an array.  The user is queried on each successive row, starting from the top, until they indicate the row which contains their target character.  If the user does not generate a response, the process restarts from the first row.  Once a row is selected, items are queried from left to right until another positive indication is received.  The block-row-item is similar, but there is an initial set of queries which select a block (a set of rows).

A query consists of the system asking the user, `Is your target character in this set?'.  One could build a system with only two input buttons corresponding to a `yes' or `no' response.  We call such a system \textit{binary} switch scanning.  Alternatively, the system can be reduced to a single `yes' input and have the user indicate `no' by allowing some fixed amount of time to pass; we refer to this as \textit{timed} switch scanning.

Row-item schemes are common among AACs.  Intel recently publicly released their ACAT system which first queries whether the target character is in the top or bottom half of the rows before row-item scanning as described above \cite{intelACAT}.  While visually quite different, Hex-o-spell is also a row-item scheme; it arranges groups of characters in a circle and scans clockwise for positive intent \cite{Blankertz2006}.  While we include a simplified version of each for comparison it is important to note that both layouts have additional objectives beyond EQPD.  In particular, Intel's ACAT allows the user to select from among the 10 most likely word completions as well as other PC control inputs.  Hex-o-spell's layout was heavily inspired by a mouse control application where it was natural to use only 5 of the 6 adjacent hexagons in the `item' selection\cite{Williamson2006}.

Block-row-item and row-item schemes are well studied, including incorporating word prediction \cite{Koester1994, intelACAT}, leveraging user errors to adaptively change the scan delay \cite{Simpson1999} and extending the paradigm to an arbitrary number of dimensions (e.g. block1-block2-block3-row-item) \cite{Baljko2006}.  

Row-item style keyboards have been evaluated on multiple performance models which explicitly account for human behavior via Fitts and Hick-Hyman's laws \cite{Baljko2006, Bhattacharya2008model}.  Further, multiple works compare different row-item style keyboard arrangements \cite{Lesher1998, Venkatagiri1999, MacKenzie2012}.  In the works most similar to our own, Baljko and Roark design decision trees based on Huffman codes to offer a minimum number of trials per decision \cite{Baljko2006, RoarkConf, RoarkAAC2015}; the major distinction is that our formulation considers that some trials take more time than others (selection of the first query set takes less time that selection of the second query set).

We seek a decision tree which allows users to spell faster.  To do so, we minimize the expected queries per decision, EQPD, by relaxing the row-item or similar style constraint and pose the problem as a Huffman coding variant (Karp) and import its known solution \cite{Golin1995}.  This solution explicitly searches over all decision trees.  Further, we model the user at the query level by their probability of detection and false alarm and extrapolate performance characteristics at the character level.  Finally, we perform switch scanning experiments, both timed and binary, with various layouts to highlight the relative performance benefits of each.

\section{Notation and Vocabulary}\label{sec:notationAndVocab}
We use the term \textit{query} to denote the presentation of some set of characters to the user.  A \textit{trial} describes the set of queries which partitions the set of currently viable characters (those which the user hasn't selected against).  Finally, a \textit{decision sequence} is a set of trials which selects a single character from the set of all possible characters. 

We use binary random variable $T$ to denote the target state of a query where $T=1$ indicates that the user's target character is present.  $X$ indicates the target state of a trial where $X=x_i$ indicates that the user's target character is present in the i-th query of a trial.  We use discrete random variable $M$ to denote the target character.  In keeping consistent with Information Theory literature, we denote the sequence of trials required to select $m_i$ as the character's \textit{codeword} and express it as $c_i$.  We use the hat notation to indicate any estimate.  See Fig \ref{fig:fig1} for an example decision sequence which illustrates all of these notations.

\section{Row-Item: Minimizing EQPD}\label{sec:rowcol}
We define the expected codeword cost as:
\begin{align*}
E[\text{cost}(c_i)] &= \sum_i P_M(m_i) \text{cost}(c_i) \\
 &= \sum_i P_M(m_i) \sum_{x_i \in c_i} \text{cost}(x_i) \numberthis
\label{eqn:obj}
\end{align*}
where $x_i \in c_i$ refers to all trials within a codeword (counting repeated trials more than once).  Assuming that each query takes the same amount of time and noting that it takes $i$ queries to select $x_i$, we define a time based cost as:
\begin{equation}
\label{eqn:timeCost}
\text{cost}_{t}(x_i) = i
\end{equation}
With this cost, minimizing (\ref{eqn:obj}) amounts to minimizing EQPD.  For example, in a row-item scheme the number of queries required to decide on a character is based on that character's position within the array:
\begin{table}[H]
\centering
\begin{tabular}{|l|l|l|l|l|l|}
\hline
2 & 3 & 4 & 5 & 6  & 7  \\ \hline
3 & 4 & 5 & 6 & 7  & 8  \\ \hline
4 & \textcolor{red}{5} & 6 & 7 & 8  & 9  \\ \hline
5 & 6 & 7 & 8 & 9  & 10 \\ \hline
6 & 7 & 8 & 9 & 10 & 11  \\ \hline
\end{tabular}
\caption{$5 \times 6$ row-item scheme costs, $\text{cost}_t(c_i)$, by array position corresponding to Fig \ref{fig:fig1}. \label{tbl:rowColCost}} 
\end{table} 
such that to select letter `N' in the 3rd row, 2nd item of Fig \ref{fig:fig1}:
\begin{equation}
\text{cost}_t(c_i) = \text{cost}_t(x_3) + \text{cost}_t(x_2) = 3 + 2 = 5
\label{eqn:exCost}
\end{equation}
Note that this objective does not consider the initial time it takes a user to find their target character as the Hick-Hyman law (see \cite{fittsHickHyman2005}) would.  However, most scanning keyboards fix the position of the characters.  Therefore, adding the initial selection time to our objective would produce a constant delay to each decision sequence.  Because of this fact, a Hick-Hyman extension to this objective would not change the optimal decision tree.

To minimize EQPD in row-item layouts, place characters in the array such that more likely characters are assigned to lower costs.  Generally, this will yield an upper-left triangular row-item scheme which leaves the rightmost elements of each row blank \cite{Foulds1975}.  

Even so, any row-item scheme necessarily has two trials per letter selection,  might we be able to reduce (\ref{eqn:obj}) further by first querying whether the target row is in the top or bottom half of all rows, as is done in Intel's ACAT system \cite{intelACAT}?  Moreover, must all codewords have the same number of trials?  In the following section, we broaden our solution space beyond row-item schemes by reformulating our problem as a Huffman code with unequal symbol costs. 

\section{Decision Trees}\label{sec:dTrees}
\begin{figure}
\centering
\includegraphics[width=0.35\textwidth]{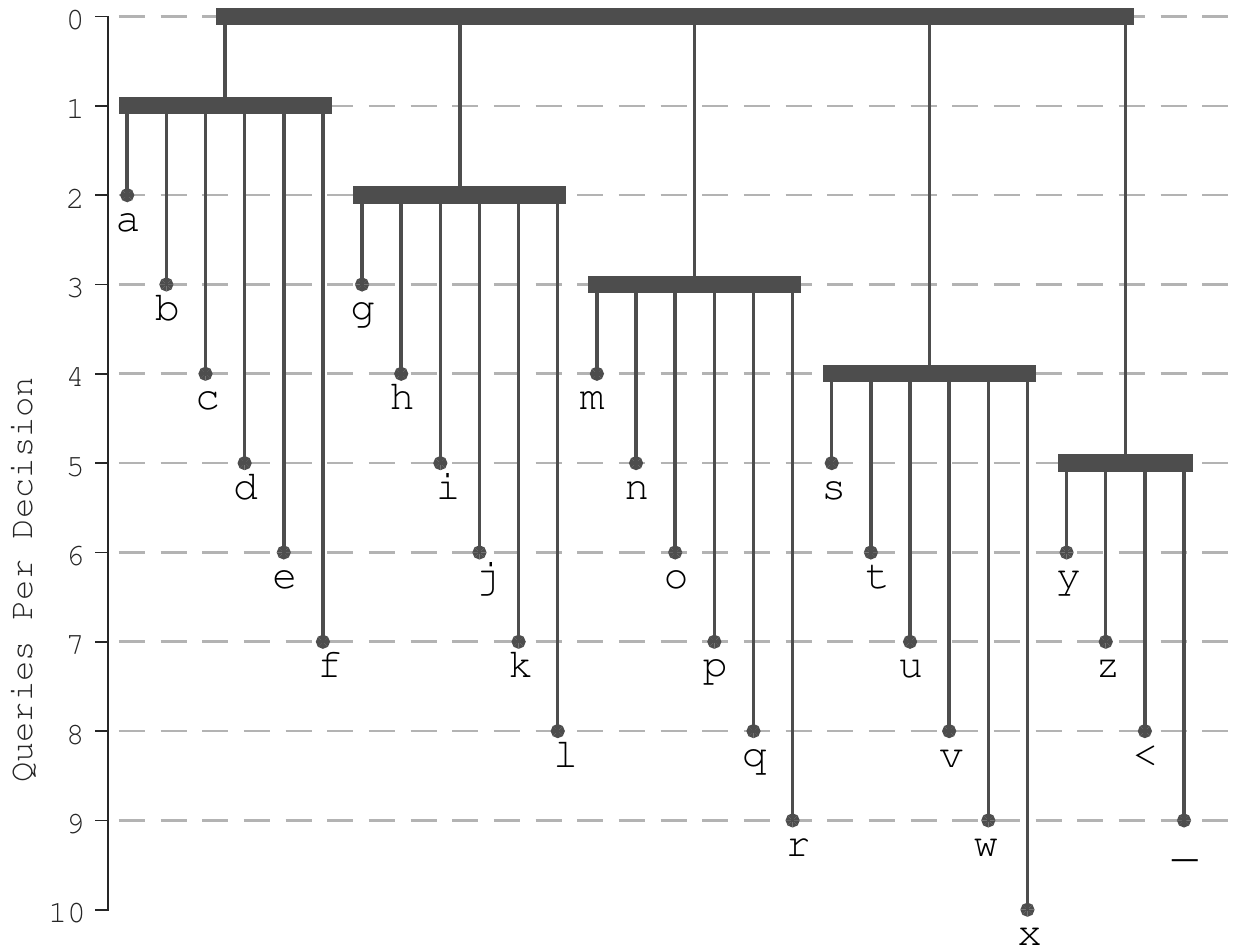}
\caption{Equivilant decision tree of the $5 \times 6$ alphabetical row-item scheme of Fig \ref{fig:fig1}. \label{fig:fig2}}
\end{figure}
It is intuitive to consider any decision scheme as a tree (see Fig \ref{fig:fig2}).  Each trial is represented by a node, drawn as a horizontal bar.  Any decision starts at the root trial, the highest horizontal bar.  At every trial node, the user is queried as to which of the current node's children their target character belongs to.  The current node is then updated according to the user input and the process progresses downwards until a unique character is reached and a decision is made; see our video \footnote{\url{https://www.youtube.com/watch?v=3LUxpeVNE2U}} for an example.  

Downward edges are drawn to scale with how many queries are required to move to a particular node, as in (\ref{eqn:timeCost}).  Returning to the same `N' selection example, Fig \ref{fig:fig2} shows that we must move along an edge associated with 3 queries and then another edge associated with 2 queries (see also Fig \ref{fig:fig1} or (\ref{eqn:exCost})).  

Note that there can only be a single i-th query; any node's outward edge costs must be unique positive integers.  Later, it will be significant that a trial's target state may take any value in the infinite set $\{x_1, x_2, \hdots \}$.

In addition to our objective (\ref{eqn:obj}), we also require that we can uniquely decode a received codeword.  For example, if $c_1 = [x_4]$ and $c_2 = [x_4, x_4]$ then when the system receives input $x_4, x_4$ there is ambiguity as to whether the user is indicating $m_1, m_1$ or $m_2$.  For an encoding function to be uniquely decodable it is sufficient to require that all codewords be prefix free, where no codeword is the prefix of another codeword.  It is encouraging to know that if we already need a uniquely decodable code we lose nothing, in terms of our objective, by further requiring it be prefix free code; see the Kraft inequality in \cite{MacKay2003, cover2012elements}.

Fortunately, codeword sets which can be represented as a tree are prefix free; no character is the descendant of another.  Similarly, it is not difficult to construct a tree from a set of prefix free codewords.  Given this bijection, we shift our search for a prefix free code which minimizes (\ref{eqn:obj}) to a tree which does the same.  

\section{Karp Codes}\label{sec:karp}
We seek a codeword for each character which is both prefix free and minimizes the expected codeword cost.  This problem has been called Huffman coding with unequal symbol (trial) cost or a Karp code after Richard Karp who first solved the problem \cite{Karp1961}.  Restricting ourselves to the case of integer symbol (trial) costs, Golin et al. improve on the computational cost of the solution.  Golin's solution explicitly searches the space of trees by enumerating all tree equivalence classes, sets of trees whose objective (\ref{eqn:obj}) is the same.  To do so, Golin defines a tree branch growth operation which relates equivalence classes and quantifies the operation's impact on the objective.  While full details are outside the scope of this work, we direct the interested reader to \cite{Golin1995} for theoretical detail or our MATLAB implementation\footnote{ \url{http://hdl.handle.net/2047/d20194049}}.

Golin's method requires a finite set of symbol (trial) costs.  As described in Sec \ref{sec:dTrees}, our search space has no such constraint and may use any member of $\{ x_1, x_2, \hdots \}$.  We claim that if a set of codewords is optimal over $\{ x_1, \hdots x_n \}$ with maximum codeword cost less than or equal to $n+1$ then it is also optimal over $\{ x_1, x_2, \hdots \}$.  

To see this, let $C_n$ be the set of codewords which minimize EQPD using only trial symbols in $\{x_1, \hdots x_n\}$.  Assume that $C_n$ has maximum codeword cost at most $n+1$.  Using $x_{n+1}$ in $C_n$ cannot lower EQPD; any codeword which uses $x_{n+1}$ has a cost at least as high as $n+1$, but all codewords already have cost at most $n+1$.  It follows that $C_n = C_{n+1}$.  We arrive at our claim by induction. 

Practically speaking, we first perform the optimization over $\{x_1, \hdots x_n\}$ for some $n$.  If the resultant codewords have maximum cost $n+1$ queries or less then these codewords are also optimal for the infinite set $\{x_1, x_2, \hdots \}$.  Otherwise, we increment $n$ and re-compute until the maximum codeword cost is at most $n+1$.

\begin{figure*}
    \centering
    \begin{tabular}{ccc}
    \begin{subfigure}{0.28\textwidth}
        \includegraphics[width=\textwidth]{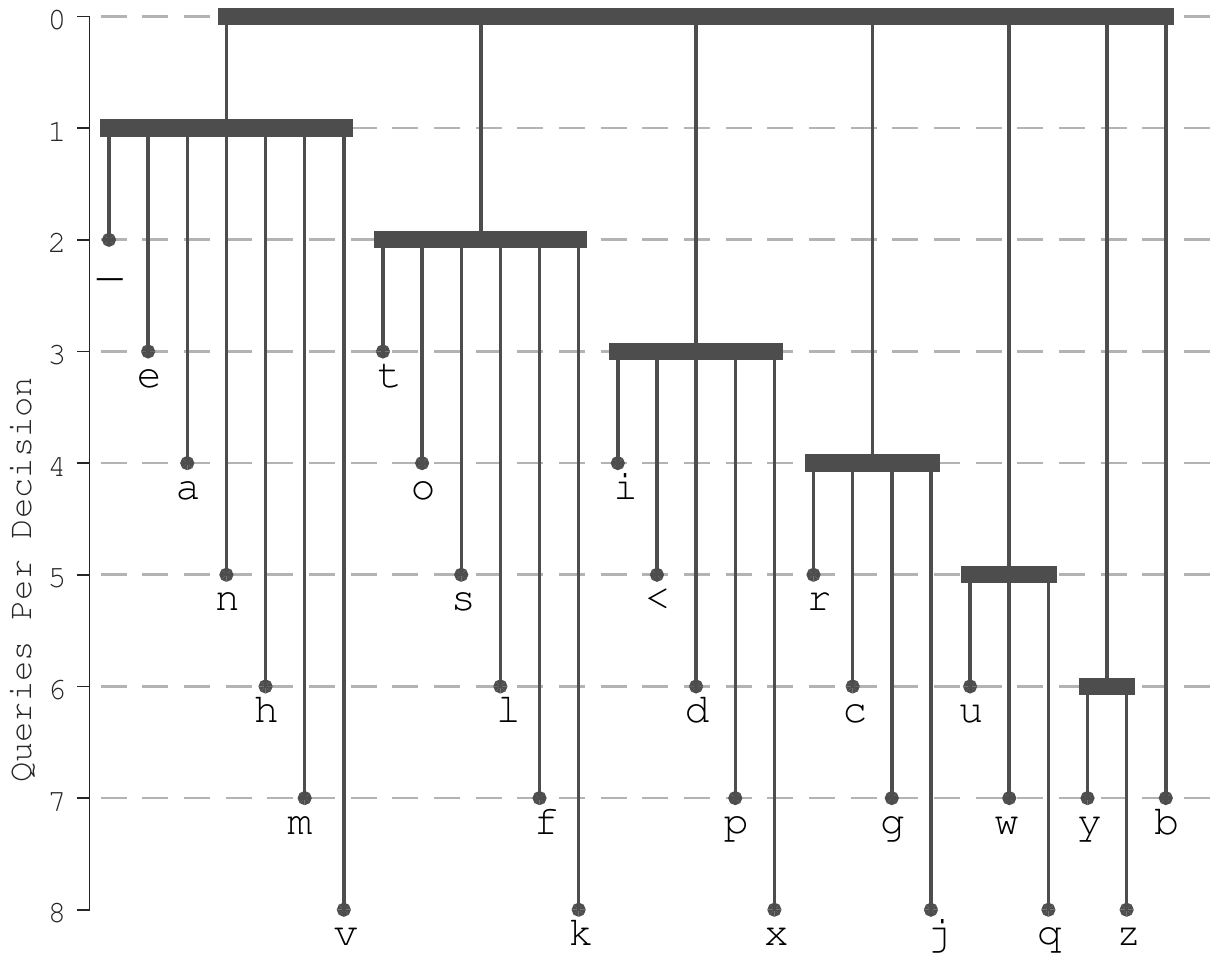}
        \caption{\textbf{Row-Item Min EQPD}\label{fig:fig5a}}
    \end{subfigure} &
    
    \begin{subfigure}{0.28\textwidth}
        \includegraphics[width=\textwidth]{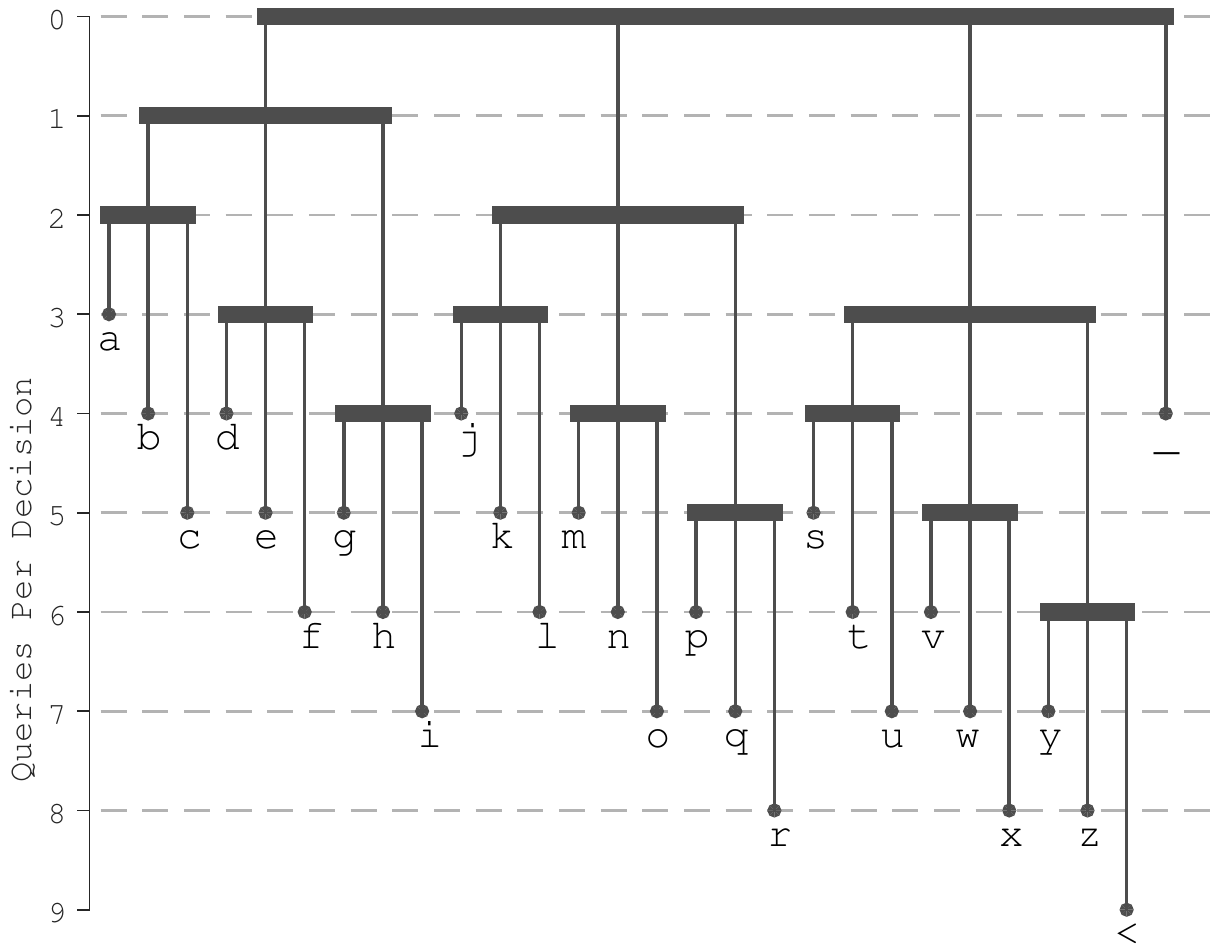}
        \caption{\textbf{Block-Row-Item Alpha} \label{fig:fig5b}}
    \end{subfigure} &
    
    \begin{subfigure}{0.28\textwidth}
        \includegraphics[width=\textwidth]{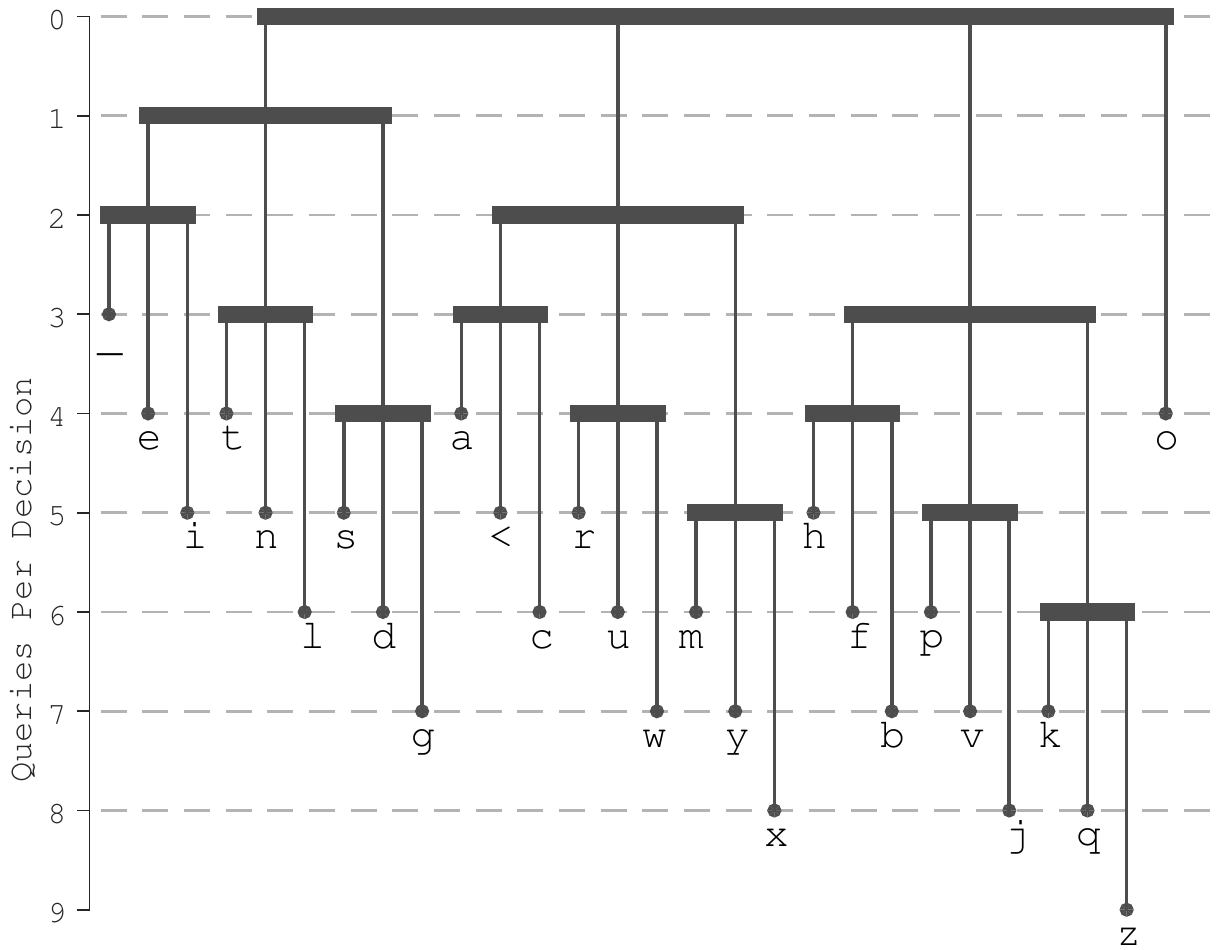}
        \caption{\textbf{Block-Row-Item Min EQPD}\label{fig:fig5c}}
    \end{subfigure} \\
    
    \begin{subfigure}{0.28\textwidth}
        \includegraphics[width=\textwidth]{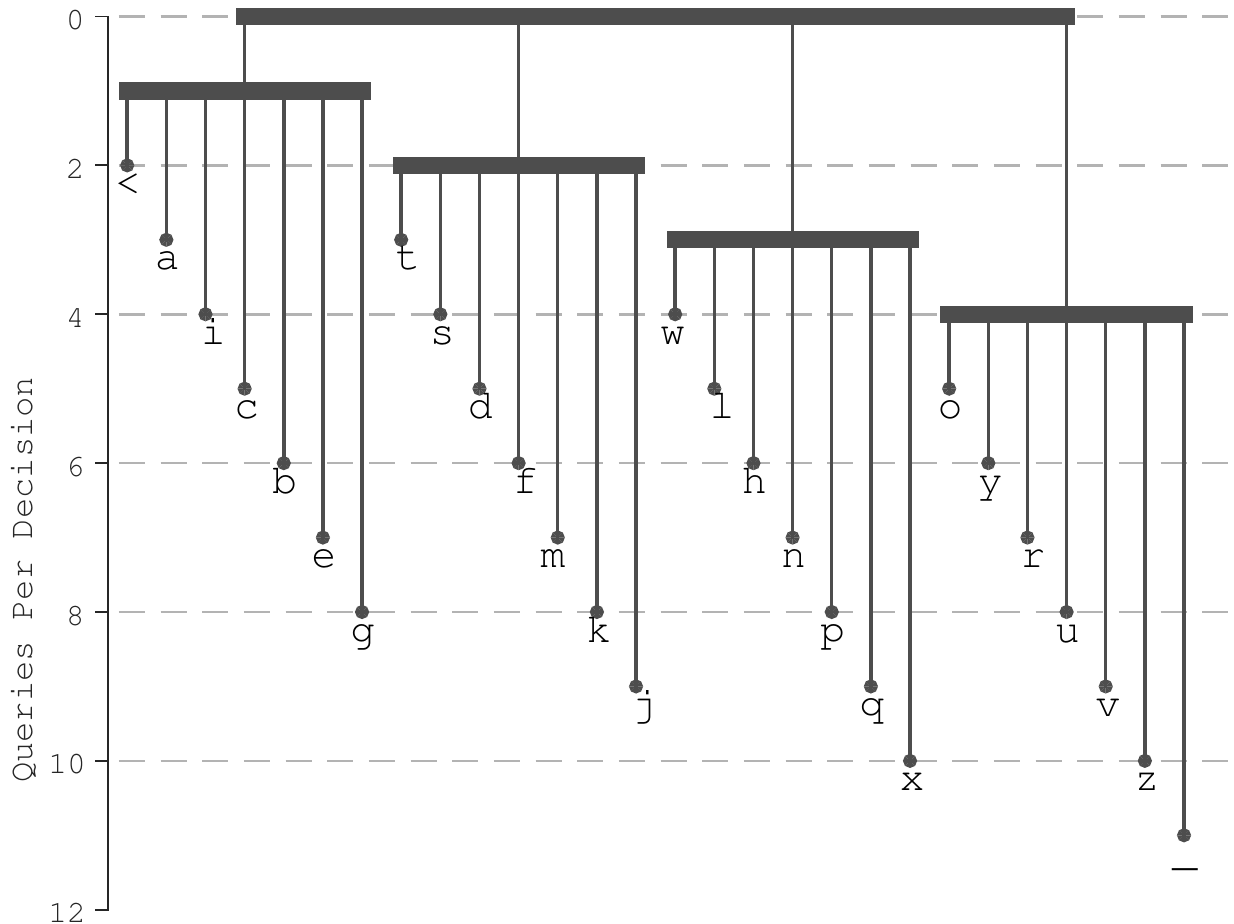}
        \caption{\textbf{Intel ACAT}\label{fig:fig5d}}
    \end{subfigure} &
    
    \begin{subfigure}{0.28\textwidth}
        \includegraphics[width=\textwidth]{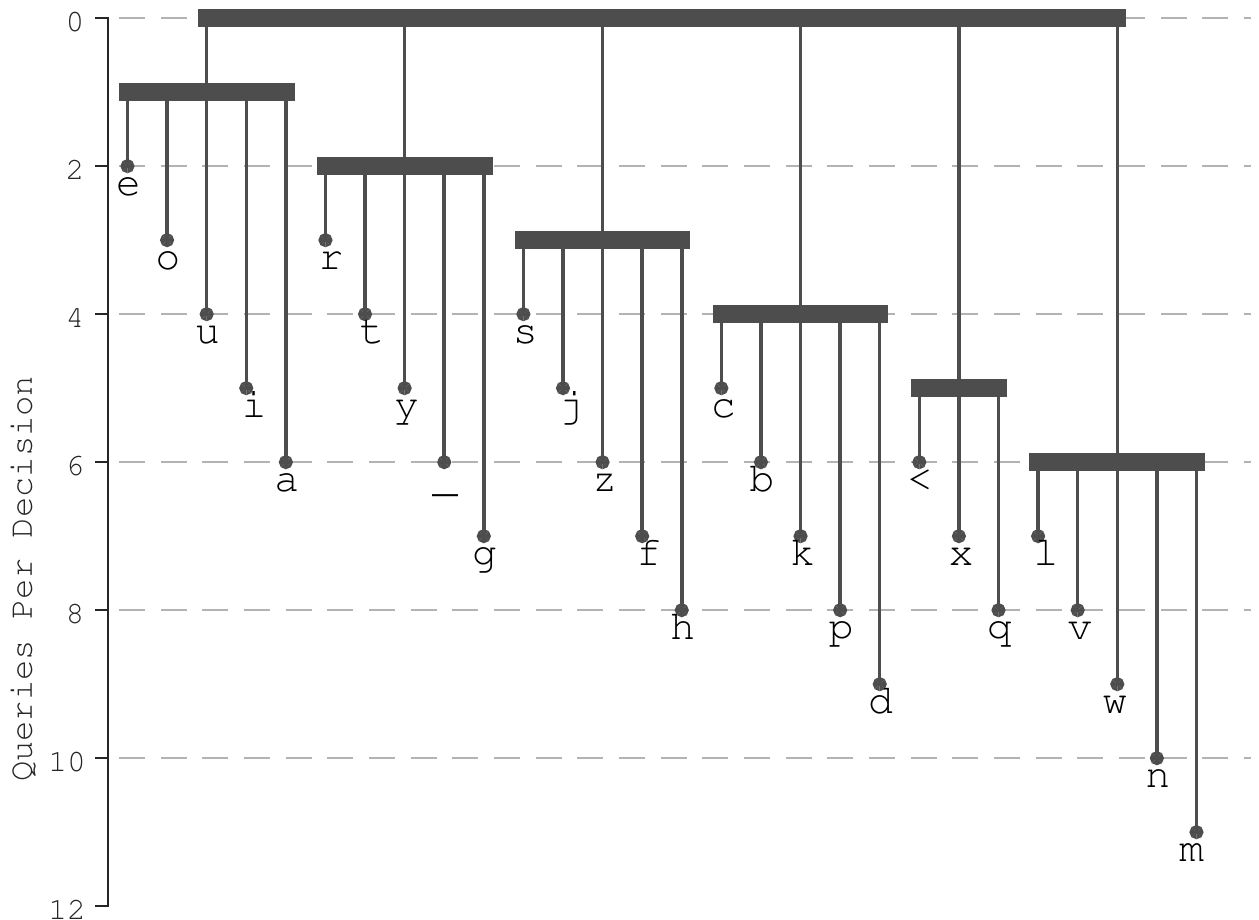}
        \caption{\textbf{Hex-o-spell}\label{fig:fig5e}}
    \end{subfigure} &
    
    \begin{subfigure}{0.28\textwidth}
        \includegraphics[width=\textwidth]{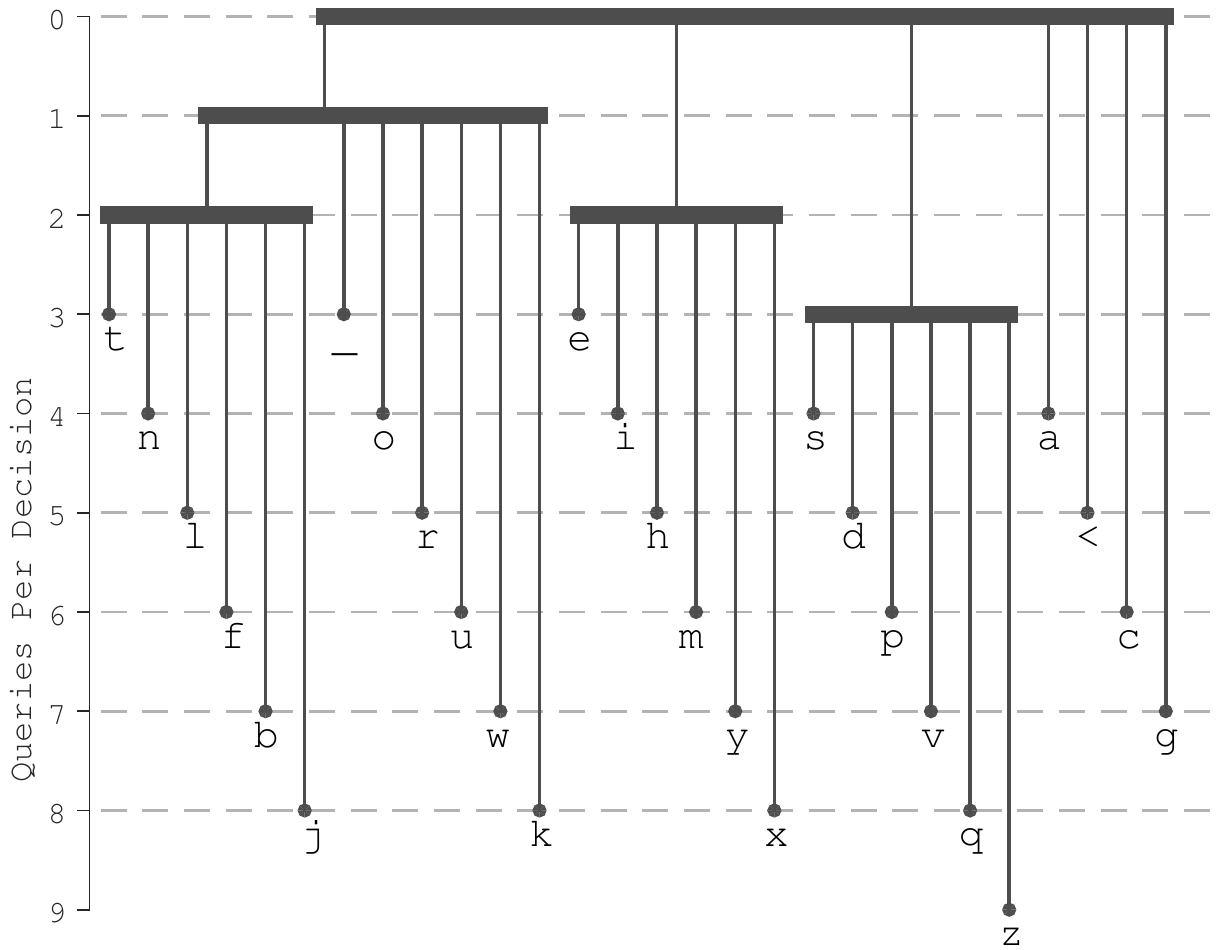}
        \caption{\textbf{Karp}\label{fig:fig5f}}
    \end{subfigure}
     \end{tabular}
    \caption{Decision Trees, see bolded descriptions in Sec \ref{sec:results} for detail.  \label{fig:otherdtrees}}
\end{figure*}

\section{Modeling Performance}\label{sec:performance}
In an effort to compare different decision trees, as well as predict user performance, we seek a model which extrapolates character performance from query level performance.  We assume that successive queries are independent such that we can parametrize a user's query behavior by their probability of false alarm, $P_{\hat{T}|T}(1|0)$, and probability of detection, $P_{\hat{T}|T}(1|1)$.  We can then compute the probability that a user selects their target character as:
\begin{equation}
P_{\hat{X}|X}(x_i|x_i)^* = P_{\hat{T}|T}(0|0)^{i-1} P_{\hat{T}|T}(1|1)
\label{eqn:trialAccFirstTime}
\end{equation}
because it requires $i-1$ no intents and a single positive intent to select $X_i$.  However, it is possible that the user shows no intent through the entire trial, at which point we assume that the trial repeats itself.  Because of this, the above probability only accounts for selection upon first seeing the target character (hence the asterisk).  More generally, 
\begin{align*}
P_{\hat{X}|X}(x_i|x_i) &= \sum_{k=0}^\infty P_{\hat{T}|T}(0|0)^{(i-1)+k(N-1)} P_{\hat{T}|T}(0|1)^{k} P_{\hat{T}|T}(1|1) \\
&= P_{\hat{T}|T}(0|0)^{(i-1)} P_{\hat{T}|T}(1|1)   \hdots \\ 
& \quad \quad \quad \sum_{k=0}^\infty [P_{\hat{T}|T}(0|0)^{N-1} P_{\hat{T}|T}(0|1)]^k \\
&= \frac{P_{\hat{T}|T}(0|0)^{(i-1)} P_{\hat{T}|T}(1|1)}{1-P_{\hat{T}|T}(0|0)^{N-1} P_{\hat{T}|T}(0|1)} \numberthis
\label{eqn:trialAcc}
\end{align*}
where $k$ denotes how many times the target character is passed in a trial, $N$ is the number of queries in the trial and we make use of the geometric series expression:
\begin{equation}
a\sum_{k=0}^\infty r^k = \frac{a}{1-r}
\label{eqn:geoSeries}
\end{equation}
in the final equality.  We can then compute character accuracy:
\begin{equation}
P_{\hat{M}|M}(m_i|m_i) = \prod_{x_j \in c_i} P_{\hat{X}|X}(x_j|x_j)
\label{eqn:srcMsgAcc}
\end{equation}
where we use $x_j$ to denote the target in trial $j$ required to select $m_i$.  Finally, the expected character accuracy is:
\begin{equation}
P(\hat{M} = M) = \sum_i P_{\hat{M}|M}(m_i|m_i) P_M(m_i)
\label{eqn:taskSymbolAcc}
\end{equation}

\section{Results}\label{sec:results}
Each of the decision trees was simulated by the model developed in Sec \ref{sec:performance} and tested by 10 users in \textit{timed} and \textit{binary} modes.  In order to offer a fair comparison, we fix each decision tree so that none can adapt to local context.  This has the added benefit of lowering the cognitive load on the user as they can eventually memorize the static layout.  For this reason, we use the average letter frequency in the English language (see Fig \ref{fig:fig3}) rather than an n-gram style approach which would vary with context.  We have deduced the probability of the space character `\texttt{\_}' based on an average word length of 4.79 and re-normalized to allow for a backspace probability of $5\%$ \cite{nGramNorvig}.  We build the following decision trees:

\begin{figure}
\centering
\includegraphics[width=0.4\textwidth]{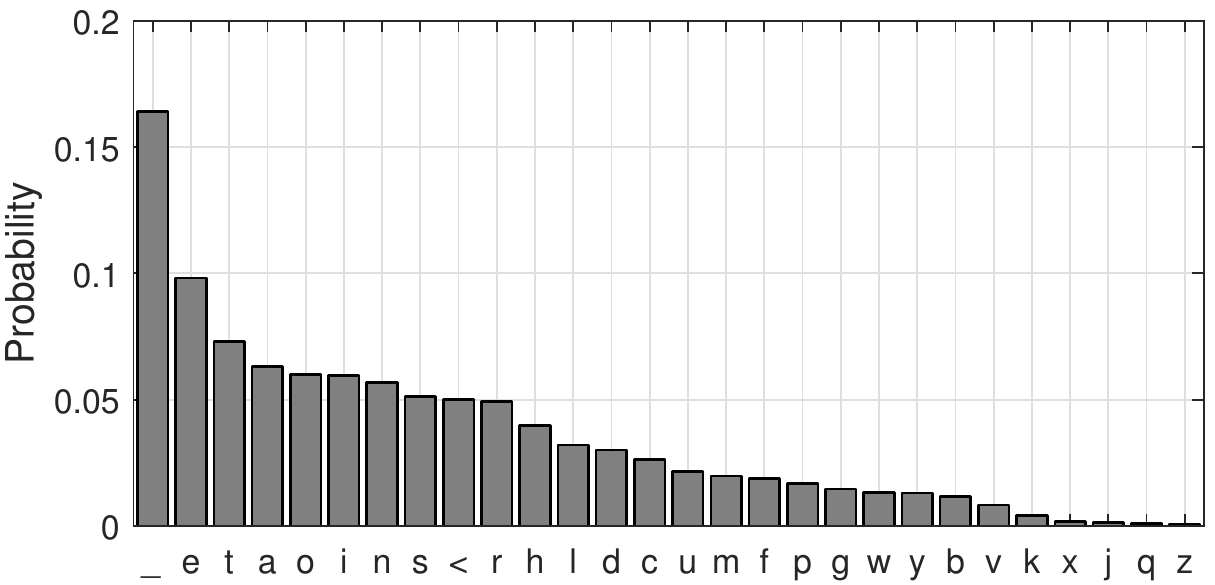}
\caption{Letter Frequencies from \cite{nGramNorvig}.  `\texttt{\_}' and `$<$' indicate space and backspace respectively. \label{fig:fig3}}
\end{figure}

\begin{enumerate}
\item \textbf{Row-Item Alphabetical}: Letters placed alphabetically in a $5 \times 6$ array, see Figs \ref{fig:fig1} and \ref{fig:fig2}.
\item \textbf{Row-Item Min EQPD}: Minimize EQPD constrained to Row-Item layouts of arbitrary size, see Fig \ref{fig:fig5a}.
\item \textbf{Block-Row-Item Alphabetical}: Letters placed in an approximately $3 \times 3 \times 3$ array alphabetically, see Fig \ref{fig:fig5b}.
\item \textbf{Block-Row-Item Min EQPD}: Letters placed in an approximately $ 3 \times 3 \times 3$ array to minimize the EQPD, see Fig \ref{fig:fig5c}.
\item \textbf{Intel ACAT} Fig \ref{fig:fig5d} and \cite{intelACAT}
\item \textbf{Hex-o-spell} Fig \ref{fig:fig5e} and \cite{Blankertz2006}
\item \textbf{Karp}: Minimize EQPD, see Sec \ref{sec:karp} and Fig \ref{fig:fig5f}.
\end{enumerate}

The Karp algorithm does not enforce any physical layout of characters which ensures that query sets are contiguous, though it is not difficult to create one manually (see Fig \ref{fig:fig7}).  The behavior of this layout may not be as intuitive as a row-item layout.  In particular, the first trial queries the first two rows together, the third row, the fourth row and then item by item within the fifth row.  Standard row-item layouts allow a user to build a strong intuition of the next query set, allowing them to think ahead.  Considerations against complexity are not part of the Karp tree's design.  We examine the significance of these human elements in Sec \ref{ssec:expResults}.

\subsection{Simulated Results}\label{ssec:simResults}
\begin{figure}
\centering
\includegraphics[width=0.35\textwidth]{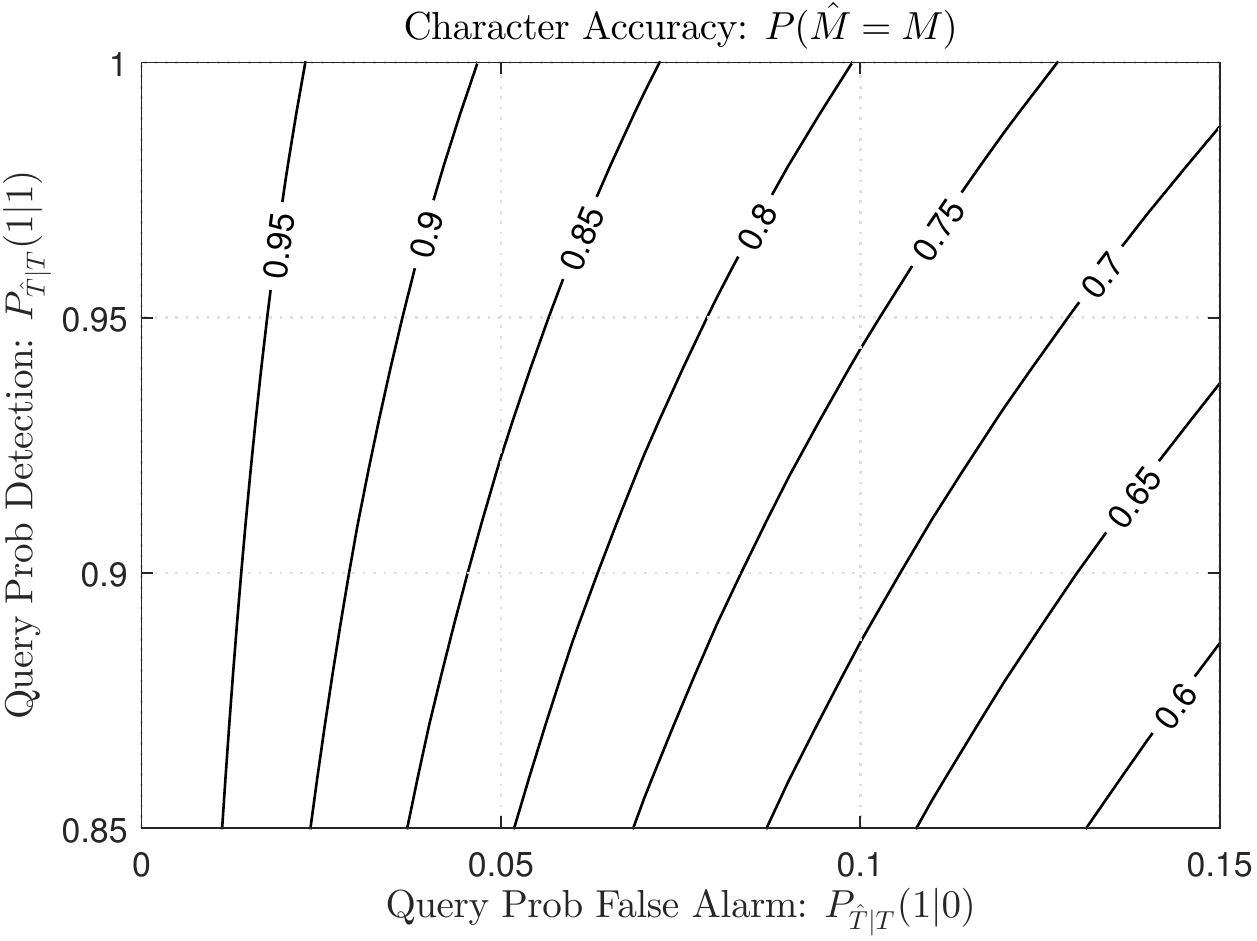}
\caption{Letter selection accuracy (\ref{eqn:taskSymbolAcc}) of the Karp decision tree (Fig \ref{fig:fig5f}) across various query level probability of detection, $P_{\hat{T}|T}(1|1)$, and probability of false alarm, $P_{\hat{T}|T}(1|0)$.  \label{fig:fig6}}
\end{figure}

Additionally, we plot the expected performance across various query level probability of detection and false alarm rates (Fig \ref{fig:fig6}).  Such a prediction would allow us to estimate a user's character level performance before they use a given layout.  Comparing our character level predictions with actual performance (see next section) yields a mean squared error of $.0048$ across all users, keyboards and modes.  The model was not more or less accurate for a particular mode.

\begin{figure}
\centering
\includegraphics[width=0.15\textwidth]{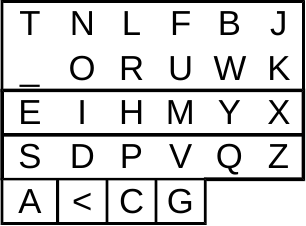}
\caption{Physical layout of optimal Karp code in Fig \ref{fig:fig5f}.  Boxes are drawn around query sets of the first trial. \label{fig:fig7}}
\end{figure}

\subsection{Experimental Results}\label{ssec:expResults}

\begin{table*}[ht]
\centering
\begin{tabular}{lc|ccc|ccc|}
\cline{3-8}
                                 & \multicolumn{1}{l|}{}              & \multicolumn{3}{c|}{\textbf{Timed Decisions}}                                                                                                                                                                        & \multicolumn{3}{c|}{\textbf{Binary Decisions}}                                                                                                                                                                      \\ \cline{3-8} 
                                 & \multicolumn{1}{l|}{\textbf{EQPD}} & \textbf{\begin{tabular}[c]{@{}c@{}}Accuracy \\ (\%)\end{tabular}} & \textbf{\begin{tabular}[c]{@{}c@{}}Time\\ (sec)\end{tabular}} & \textbf{\begin{tabular}[c]{@{}c@{}}Rollover Corrected\\ Time (sec)\end{tabular}} & \textbf{\begin{tabular}[c]{@{}c@{}}Accuracy\\ (\%)\end{tabular}} & \textbf{\begin{tabular}[c]{@{}c@{}}Time\\ (sec)\end{tabular}} & \textbf{\begin{tabular}[c]{@{}c@{}}Rollover Corrected\\ Time (sec)\end{tabular}} \\
\rowcolor[HTML]{C0C0C0} 
\textbf{Row-Item 5 x 6 Alpha}    & 6.23                               & 99                                                                & 5.60                                                          & 5.53                                                                             & 95                                                               & 1.56                                                          & 1.43                                                                             \\
\textbf{Row-Item Min EQPD}       & 4.41                               & 99                                                                & 3.50                                                          & 3.40                                                                             & 94                                                               & 1.48                                                          & 1.39                                                                             \\
\rowcolor[HTML]{C0C0C0} 
\textbf{Block-Row-Item Alpha}    & 5.60                               & 99                                                                & 4.66                                                          & 4.60                                                                             & 92                                                               & 2.05                                                          & 1.84                                                                             \\
\textbf{Block-Row-Item Min EQPD} & 4.69                               & 99                                                                & 3.42                                                          & 3.38                                                                             & 90                                                               & 2.04                                                          & 1.90                                                                             \\
\rowcolor[HTML]{C0C0C0} 
\textbf{Intel ACAT}              & 6.20                               & 99                                                                & 4.97                                                          & 4.97                                                                             & 95                                                               & 1.51                                                          & 1.41                                                                             \\
\textbf{BBCI Hex-o-spell}        & 5.55                               & 99                                                                & 4.08                                                          & 4.00                                                                             & 95                                                               & 1.33                                                          & 1.25                                                                             \\
\rowcolor[HTML]{C0C0C0} 
\textbf{Karp}                    & 4.29                               & 98                                                                & 3.28                                                          & 2.98                                                                             & 92                                                               & 1.99                                                          & 1.79                                                                             \\ \cline{3-8} 
\end{tabular}
\caption{Comparison of performance across different keyboard layouts.  EQPD was computed via (\ref{eqn:obj}).  All times are averaged across all sequences of all users after discarding incorrect sequences.  Accuracy is averaged similarly except we have included incorrect sequences.\label{tbl:results}}
\end{table*}

Ten neurotypical volunteers used each of the keyboard layouts in both \textit{timed} and \textit{binary} modes.  The volunteers were aged 19-33, all with normal or corrected to normal vision.  Each user gave written informed consent according to Northeastern's IRB protocol and was paid for their participation.

The scanning keyboards were implemented in MATLAB with Psychtoolbox \cite{Pelli1997} having the users press right shift to indicate $T=1$ and either left shift or wait 1.2 seconds to indicate $T=0$ in \textit{binary} and \textit{timed} modes respectively.  A video showing target selections for each keyboard is available\footnote{\url{https://www.youtube.com/watch?v=3LUxpeVNE2U}}.  The keyboards were presented to the user in a random order.  Users were asked to type 5 initial warm-up decision sequences and then 40 \textit{binary} and 20 \textit{timed} decision sequences on each keyboard.  The purpose of the warm-up decisions was to ensure that users understood how to use the keyboard.  All targets were sampled from the letter distribution given in Fig \ref{fig:fig3} to simulate the typing of actual words.  This set of letters was held constant among different layouts to account for potential sampling biases in favor of one keyboard or another.  In an effort to simulate the fact that long term users would have a strong expectation of the physical location of a character in each layout (where new users would only have an intuition in the alphabetical layouts), target letters were indicated by being colored red.

One user's \textit{timed} data was discarded as their accuracy was a significant outlier ($70\%$ for one keyboard where no other user-keyboard pair was lower than $90\%$, averaging $98\%$ across all keyboards and users).  We suspect the user did not understand instructions for \textit{timed} mode.

Table \ref{tbl:results} summarizes experimental results.  Accuracy was fairly consistent across all keyboards in either \textit{binary} or \textit{timed} modes.

Additionally, we accounted for the presence of \textit{rollovers}.  We define a \textit{rollover} as the event where a user initially responds `no' to their target query set, responds `no' to all other query sets and then `yes' upon seeing their target query set a second time.  We hypothesize that experienced users would have little to no rollovers given their expertise with a layout.

In \textit{timed} mode the Karp layout spent the most time in \textit{rollovers}.  Anecdotally, users mentioned that this layout was more confusing than the others as the next query set was difficult to predict.  The next query set was not necessarily the next row or item as it often was in other layouts.  Despite this, the Karp layout was still faster than all other layouts.  Under the assumption that each query takes the same amount of time, we would expect that the expected time would be some linear multiplier, the average time per query, of the EQPD.  Of course, not all queries take the same amount of time, even in \textit{timed} mode.  Any query where $T=1$ lasts until the user enters their positive intent, not waiting the full 1.2 seconds.  Every row-item layout, including ACAT and Hex-o-spell, have 2 positive queries per decision, every block-row-item layout has 3 positive queries and Karp has 2.04 positive queries on average.  More positive queries yields lower decision times.  This explains how Block-Row-Item Min EQPD was faster than Row-Item Min EQPD despite having a higher EQPD.

In \textit{binary} mode users were much faster than \textit{timed} mode at the expense of accuracy.  Both accuracy and time figures show a preference for row-item layouts rather than block-row-item or Karp style layouts.  We suspect this is a result of allowing the user to build an intuition about the next query set.  Of course, all users were new to switch scanning, we suspect that given time to practice the time per \textit{binary} decision would reflect the EQPD of each layout (maybe more so than \textit{timed} decisions as there will be no preference for positive intents).

\section{Conclusion}\label{sec:conclusion}
We have posed the layout of a switch scanning keyboard as a Huffman coding problem of unequal, integer symbol cost (a Karp code).  In doing so, we are able to provide the decision tree which minimizes the EQPD for a given prior distribution over characters.  Though we have opted to use the typing task because of its familiarity, it is straightforward to construct Karp codes for arbitrary AAC tasks (e.g. hyperlink selection in web navigation or wheelchair movement directions) so long as we are given some prior distribution to describe the frequency of a task message's occurrence.  Additionally, we offer a model which predicts character level performance given a user's query level probability of detection and false alarm.

However, the resulting Karp layout may be less intuitive than the reliable row by row querying of row-item style scanning.  This fact hindered the Karp tree's performance, especially in \textit{binary} switch scanning decisions.

Having a minimum EQPD was more helpful in \textit{timed} switch scanning where the Karp layout offered a significant performance boost.  Part of its success was unintentional, as it benefited from having more positive intent queries which are necessarily shorter than negative intent queries for \textit{timed} switch scanning.  As a direction of future work, we intend to assign a different cost to a positive and negative query.

We suggest that the Karp layout is a valuable lower bound for EQPD to keep in mind for AAC design.  Despite its less intuitive layout, our experiments show that the Karp layout outperformed all others in \textit{timed} switch scanning.  Moreover, real AAC users become experts in their layouts and we would expect them to be more robust to the nuances of a Karp layout potentially making it viable in \textit{binary} mode as well.
\bibliographystyle{IEEEtran}
\bibliography{refs}


\end{document}